# UN Handbook on Privacy-Preserving Computation Techniques

# Foreword

## The Task Team

The Privacy Preserving Techniques Task Team (PPTTT) is advising the UN Global Working Group (GWG) on Big Data on developing the data policy framework for governance and information management of the global platform, specifically around supporting privacy preserving techniques.

This task team is developing and proposing principles, policies, and open standards for encryption within the UN Global Platform to cover the ethical use of data and the methods and procedures for the collection, processing, storage and presentation of data taking full account of data privacy, confidentiality and security issues. Using these open standards, algorithms, policies, and principles will reduce the risks associated with handling proprietary and sensitive information.

The first deliverable of the task team is this UN Handbook on Privacy Preserving Techniques.

The terms of reference for the task team can be found at https://docs.google.com/document/d/1zrm2XpGTagVu4O1ZwoMZy68AufxgNyxI4FBsyUyHLck/.

The most current deliverables from the task team can be found in the UN Global Platform Marketplace (https://marketplace.officialstatistics.org/technologies-and-techniques).

## Members of the Task Team

| Name | Role | Organisation |
| --- | --- | --- |
| Mark Craddock | Chair | UN Global Platform |
| David W. Archer | Co-Editor | Galois, Inc. |
| Dan Bogdanov | Co-Editor | Cybernetica AS |
| Adria Gascon | Contributor | Alan Turing Institute |
| Borja de Balle Pigem | Contributor | Amazon Research |
| Kim Laine | Contributor | Microsoft Research |

| Andrew Trask | Contributor | University of Oxford \| OpenMined |
| Mariana Raykova | Contributor | Google |
| Matjaz Jug | Contributor | CBS |
| Robert McLellan | Contributor | STATSCAN |
| Ronald Jansen | Contributor | Statistics Division \| Department of Economic and Social Affairs. United Nations |
| Olga Ohrimenko | Contributor | |
| Simon Wardley | Contributor | Leading Edge Forum |
| Kristin Lauter | Reviewer | Microsoft Research |
| Nigel Smart | Reviewer | UK Leuven |
| Aalekh Sharan | Reviewer | NITI (National Institution for Transforming India), Government of India |
| Ira Saxena | Reviewer | NITI (National Institution for Transforming India), Government of India |
| Rebecca N. Wright | Reviewer | Barnard College |
| Eddie Garcia | Reviewer | Cloudera |
| Andy Wall | Reviewer | Office for National Statistics |

# Executive Summary

An emerging reality for statistical scientists is that the cost of data collection for analysis projects is often too high to justify those projects. Thus many statistical analysis projects increasingly use administrative data – data gathered by administrative agencies in the course of regular operations. In many cases, such administrative data includes content that can be repurposed to support a variety of statistical analyses. However, such data is often sensitive, including details about individuals or organizations that can be used to identify them, localize their whereabouts, and draw conclusions about their behavior, health, and political and social agendas. In the wrong hands, such data can be used to cause social, economic, or physical harm.

Privacy-preserving computation technologies have emerged in recent years to provide some protection against such harm while enabling valuable statistical analyses. Some kinds of privacy-preserving computation technologies allow computing on data while it remains encrypted or otherwise opaque to those performing the computation, as well as to adversaries who might seek to steal that information. Because data can remain encrypted during computation, that data can remain encrypted "end-to-end" in analytic environments, so that the data is immune to theft or misuse. However, protecting such data is only effective if we also protect against what may be learned from the output of such analysis. Additional kinds of emerging privacy-preserving computation technologies address this concern, protecting against efforts to reverse engineer the input data from the outputs of analysis.

Unfortunately, privacy-preserving computation comes at a cost: current versions of these technologies are computationally costly, rely on specialized computer hardware, are difficult to program and configure directly, or some combination of the above. Thus National Statistics Offices (NSOs) and other analytic scientists may need guidance in assessing whether the cost of such technologies can be appropriately balanced against resulting privacy benefits.

In this handbook, we define specific goals for privacy-preserving computation for public good in two salient use cases: giving NSOs access to new sources of (sensitive) Big Data; and enabling Big Data collaborations across multiple NSOs. We describe the limits of current practice in analyzing data while preserving privacy; explain emerging privacy-preserving computation techniques; and outline key challenges to bringing these technologies into mainstream use. For each technology addressed, we provide a technical overview; examples of applied uses; an explanation of modeling adversaries and security arguments that typically apply; an overview of the costs of using the technology; an explanation of the availability of the technology; and a Wardley map that illustrates the technology's readiness and suggested development focus.

## Handbook Purpose and Target Audience

This document describes motivations for privacy-preserving approaches for the statistical analysis of sensitive data; presents examples of use cases where such methods may apply; and describes relevant technical capabilities to assure privacy preservation while still allowing analysis of sensitive data. Our focus is on methods that enable protecting privacy of data *while it is being processed*, not only while it is at rest on a system or in transit between systems. This document is intended for use by statisticians and data scientists, data curators and architects, IT specialists, and security and information assurance specialists, so we explicitly avoid cryptographic technical details of the technologies we describe.

## Motivation: The Need for Privacy

In December 1890, American jurists Samuel Warren and Louis Brandeis, concerned about the privacy implications of the new "instantaneous camera", argued for protecting "all persons...from having matters which they may properly prefer to keep private, made public against their will."

Today, the dangers of having our private information stolen and used against us are everyday news. Such data may be used to identify individuals, localize their whereabouts, and draw conclusions about their behavior, health, and political and social agendas. For example, it is well known that a small set of attributes can single out an individual in a population; a small number of location data points can predict where a person can be found at a given time; and simple social analytics can reveal a sexual preference. Improper use of such localization, identification, and conclusions can lead to financial, social, and physical harm.

Criminal theft of databases of such information occur thousands of times each year worldwide. Big Data – aggregating very large collections of individual data for analytical use, often without the knowledge of the individuals described – increases the risk of data theft and misuse even more. Such large databases of diverse information are often an easy target for cyber criminals attacking from outside organizations that hold or use such data. Equally concerning is the risk of *insider threats* – individuals trusted with access to such sensitive data who turn out to be not so trustworthy.

## Unprotected Data is Vulnerable to Theft

Data is vulnerable to theft by both outsiders and insiders *at rest*, for example when stored on a server; *in transit*, for example when communicated over the Internet; and *during computation*, for example when used to compute statistics. In the past, when cyber threats were less advanced, most attention to privacy was devoted to data at rest, giving rise to technologies such as *symmetric key encryption*. Later on, when unprotected networks such as the Internet became commonplace, attention was focused on protecting data in transit, giving rise to technologies

UN Handbook on Privacy-Preserving Computation Techniques 9

such as *Transport Layer Security* (TLS*)*. More recently, the rise of long-lived cyber threats that penetrate servers worldwide gave rise to the need for protecting data during computation. We restrict our scope in this handbook to technologies that protect the privacy of data *during and after computation*, because mechanisms for protection of such data while at rest on servers and in transit between servers is a well-studied problem. We call such technologies *privacy-preserving computation*. We omit discussion of data integrity and measures that support it, for example data provenance analysis, or digital signatures on data that can be unambiguously attributed to data creators.

## Wardley Maps

This document uses Wardley Maps to explain where the privacy techniques are in the cycle of genesis through to commodity. A full explanation of Wardley Maps and how to use them for developing an ICT strategy can be found in the UN Global Platform - Handbook on Information Technology Strategy.
https://marketplace.officialstatistics.org/un-global-platform-handbook-on-information-technology-strategy

# Concepts and Setting

## Motivation for Privacy-Preserving Statistics

In order to illustrate the use of privacy-preserving computation in the context of statistics, we first present two settings where confidential data is used. These are inspired by uses of privacy-preserving computation technology by National Statistics Offices (NSOs) around the world. For both settings we discuss stakeholders, data flows, privacy goals and example use cases with their privacy goals.

### Example Setting 1: Giving NSOs access to new sources of Big Data

Figure 1 illustrates a setting where a single NSO wishes to access sensitive data. As shown at left in the figure, organisations may provide such data to NSOs as the result of direct surveys or indirectly by *scraping* data from available sources. Data about individuals may be collected and provided to NSOs by intermediaries such as telephone, credit card, or payment companies. Individual data may also come from government sources, for example, income surveys or census reports. In addition, *data aggregators* that collect and trade in such information may also provide data to NSOs. We call such individuals and organizations that provide data **Input Parties** to privacy-preserving computation.

UN Handbook on Privacy-Preserving Computation Techniques            10

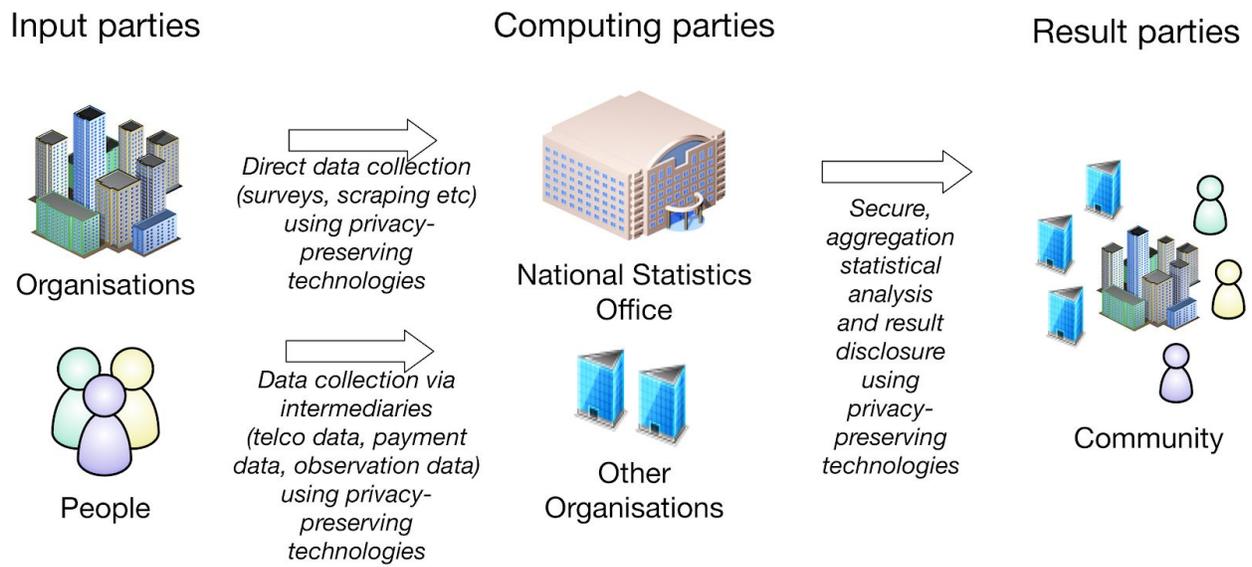

*Figure 1: Privacy-preserving statistics workflow for a single Statistics Office*

NSOs and other organizations that receive such data, shown at center in the figure, *compute* on the collected data they obtain from input parties, and thus are called **Computing Parties**. Such computation transforms the collected data into information – assemblies of data that have a specific context and structure that makes the data useful. For example, the results of such computations are often statistical reports that may be used by governments or NGOs to make decisions about the allocation of scarce resources.

Information resulting from NSO computations are then securely distributed to individuals or organizations that combine it with their existing knowledge to discover patterns that are prioritizable and actionable. We call these recipients **Result Parties**.

Throughout this simple model of data and information flow there are a multitude of *privacy risks*. We start by assuming that data is secure while it remains in the hands of the input parties – that is, we assume those parties have their own cybersecurity solutions for protecting data within their domains. Thus the first privacy risk in this setting occurs when that data is *in transit* between the input parties and computing parties. Existing technologies such as *TLS* are often used to mitigate in-transit privacy risks. The second privacy risk occurs when the data is *at rest* in the domain of the computing parties. Encryption using technologies that employ standards such as the *Advanced Encryption Standard* (AES) are often used to mitigate at-rest privacy risks. The third privacy risk in this setting occurs when the data is used for computation to produce information. In current practice, data is decrypted prior to such use. However, such decryption brings that data *into the clear*, where it may be stolen or misused. In this handbook, we focus on technologies for computing while the data remains encrypted, mitigating this privacy risk.

In addition to the risks described above, there is an at-rest privacy risk while the information resulting from computation resides with the computing party, and an in-transit privacy risk while that information is distributed to result parties. These risks are mitigated in the same way as the other at-rest and in-transit risks described above.

When result parties receive information from compute parties, privacy risks continue, because such information may still be sensitive, and may be used in some cases to infer values of input data. Additional technologies such as *differential privacy* may mitigate some or all of that risk.

**Example use case: Point-of-sale transaction data**. NSOs seek to directly collect product pricing data from multiple retailers at multiple sites to calculate econometric statistics. Retailers want to prevent their pricing data being revealed in bulk as such information might be damaging if accessed by the competition.

**Example use case: Mobile phone data.** NSOs collect cell phone location data from telecommunications operators to use in generating tourism statistics. In addition to having to protect highly sensitive data of where a person is at all times, the telecommunications operators are also liable for the protection of the data.

## Example Setting 2: Enabling Big Data Collaborations Across Multiple NSOs

Figure 2 illustrates a setting where multiple NSOs collaborate under the coordination of the United Nations. It could be said that this case is an extension of the case above. However, it differs in that individuals and organizations that provide raw data are no longer input parties. Instead, we call them **Data Subjects**, because the data of interest in this setting describes them. After collecting data as shown in the setting above and conducting statistical analysis locally, NSOs from individual nations act as Input Parties in this setting to share their results and methods with each other on the UN Global Platform. Thus in this setting, the Global Platform takes on the role of the Computing Party. Also in this setting the Result Parties may be more diverse than in the first setting above: people, organizations, and governments across the world may receive and benefit from reports produced by the Global Platform.

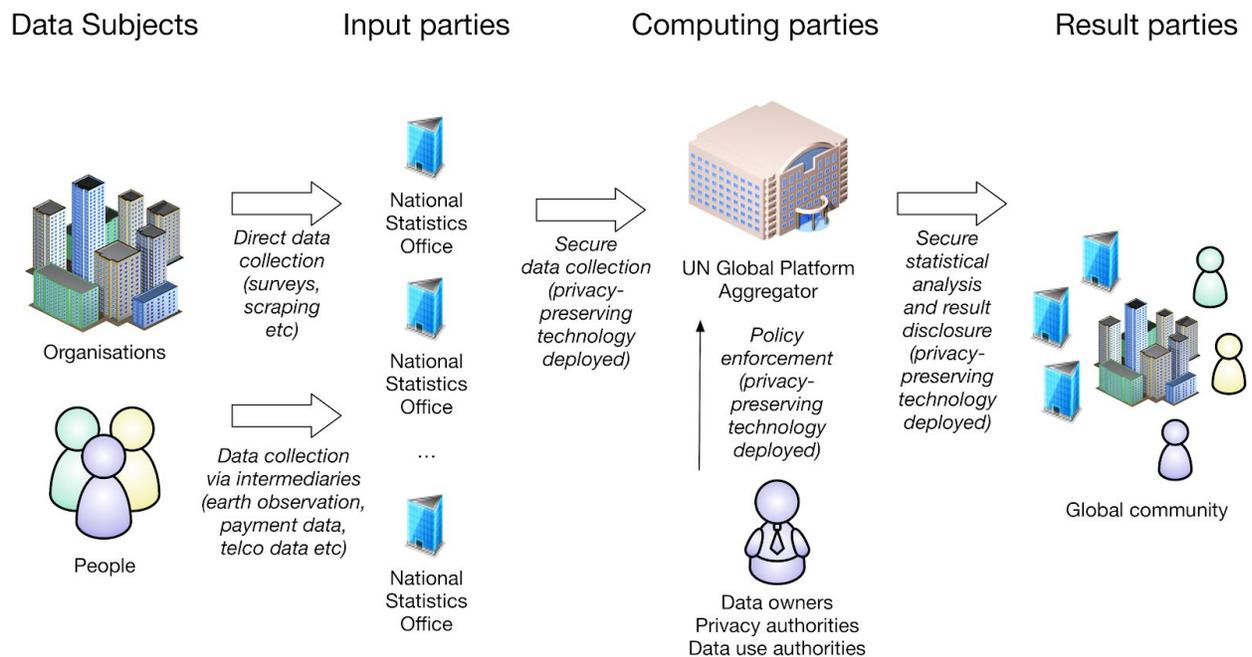

*Figure 2: Privacy-preserving statistics workflow for the UN Global Platform*

# Privacy Goals for Statistical Analysis

## Privacy Threats and the Role of Privacy Enhancing Technologies

Often in general conversation about privacy, information security practitioners use a hermetic analogy: privacy is sustained to the extent that information does not "leak" outside the protection of those authorized to access it. By that analogy, all Privacy Enhancing Techniques (PETs) discussed in this handbook partially address the general question of "how much does a data analysis leak about the sensitive part of its input dataset".

The leakage may be intentional (a hacker, curious data analyst) or unintentional (unexpected sensitive result during the analysis). In any case, Privacy Enhancing Technologies can reduce the risks for such leakage.

It is important to remark that none of the Privacy Enhancing Technologies we describe, and in fact no known technique, gives a complete solution to the privacy question, mainly because such a vaguely defined goal might have different suitable interpretations depending on the context.

For this reason, while some of the discussed technologies offer complementary and thus incomparable privacy guarantees, a fully-fledged privacy-preserving data analysis pipeline necessarily must integrate several of these technologies in a meaningful way, which in turn requires understanding the interplays of their respective privacy definitions. Such integration starts at the threat modelling stage, as privacy requirements must ultimately be set in terms of the concrete parameters of the privacy definition that applies to each technology.

## Key Aspects of Deploying Privacy Enhancing Technologies

The crucial aspect in deploying PETs is that they have to be deployed as close to the data owner as possible. The best privacy guarantees require that PETs are applied by the data owner, on premises, before releasing confidential data to third parties.

This can be explained with a simple analogy – the use of access control. Typically, organisations working with data deploy role-based access control (RBAC), that grants access to data only for authorised individuals. However, this still assumes that the organisation itself has full access to all the collected data. Thus, the organisation remains liable for all data. However, with correctly deployed Privacy Enhancing Technologies, the organisation will be able to perform its duties without full access and, therefore, with reduced liability.

## Privacy Goals for Statistics

Following the general descriptions of our two settings above, we use the abstraction below to explain privacy goals. As shown in Figure 3, one or more Input Parties provide sensitive data to one or more Computing Parties who statistically analyse it, producing results for one or more Result Parties.

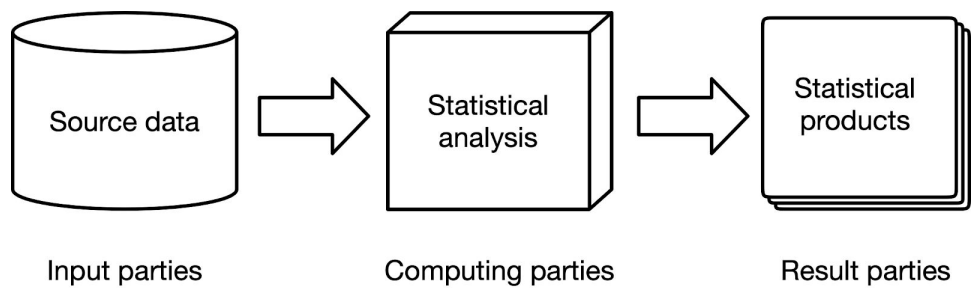

*Figure 3: Abstract setting for the privacy goals*

We now introduce three general privacy goals that naturally link to technologies and privacy definitions introduced later in the document. These goals should be regarded as a general guideline: concrete deployments are likely to have specific privacy requirements that require careful evaluation. Nevertheless, such requirements should ideally be addressed in a way that provides concrete privacy guarantees, and we see the following categorization as the natural starting point in that modelling task. The privacy goals of *input privacy*, *output privacy* and *policy enforcement* are adapted from research on privacy-preserving statistics[1],[2].

## Input Privacy

Input privacy means that the Computing Party cannot access or derive any input value provided by Input Parties, nor access intermediate values or statistical results during processing of the data (unless the value has been specifically selected for disclosure). Note that even if the Computing Party does not have direct access to the values, it may be able to derive them by using techniques such as side-channel attacks[3]. Thus input privacy requires protection against all such mechanisms that would allow derivation of inputs by the Computing Party.

Input privacy is highly desirable as it significantly reduces the number of stakeholders with full access to the input database. That, in turn, reduces liability and simplifies compliance with data protection regulations.

The notion of input privacy is particularly relevant in settings where mutually distrustful parties are involved in a computation on their private data, but where any party learning more than their prescribed output is considered a privacy breach. Referring back to the scanner data example above, the retailers would require that the system set in place to collect and calculate price indices would provide input privacy for the input prices.

## Output Privacy

A privacy-preserving statistical analysis system implements output privacy to the extent it can guarantee that the published results do not contain identifiable input data beyond what is allowable by Input Parties.

Output privacy addresses the problem of measuring and controlling the amount of leakage present in the result of a computation, regardless of whether the computation itself provides input privacy. For example, in a scenario where a distributed database provided by multiple parties is analysed to produce a statistical model of the data, output privacy has to do with the problem of how much information about the original data can be recovered from the published statistical model, but not how much information is leaked by the messages exchanged between the parties during the computation of the model, as the latter is related to input privacy.

---

[1] [K15] Liina Kamm. **Privacy-preserving statistical analysis using secure multi-party computation.** PhD thesis. University of Tartu. 2015. Available online: http://hdl.handle.net/10062/45343 (last accessed: July 2nd, 2018)

[2] [BKLS16] Dan Bogdanov, Liina Kamm, Sven Laur, Ville Sokk. **Rmind: a tool for cryptographically secure statistical analysis**. IEEE Transactions on Dependable and Secure Computing. 2016. Available online: http://dx.doi.org/10.1109/TDSC.2016.2587623 (last accessed: July 2nd, 2018)

[3] Side-channel attacks are used to derive confidential data from query timings, cache timings, power usage, electromagnetic emissions or similar measurable phenomena from the computer doing the processing.

Output privacy is highly sought after in data publication, e.g., when an NSO would like to make a database available to the general public without revealing any relevant input data used to derive the published data.

## Policy Enforcement

A privacy-preserving statistical analysis system implements policy enforcement if it has a mechanism for the input parties to exercise positive control which computations can be performed by the computing parties on sensitive inputs, and which results can be published to which result parties. Such positive control is typically expressed in a formal language that identifies participants and the rules by which they participate. Policy *decision points* process these rules into machine-usable form, while policy *enforcement points* provide technical means to assure that the rules are followed. Thus policy enforcement can describe and then automatically assure input and output privacy in a privacy-preserving statistical analysis system, thus reducing reliance on classic but less effective approaches such as non-disclosure agreements and confidentiality clauses in data use contracts.

## Combining Multiple Privacy Goals

An actual statistical system will most likely combine multiple techniques to cover multiple privacy goals. See Figure 4 for an example of how they can cover the whole system shown in Figure 3.

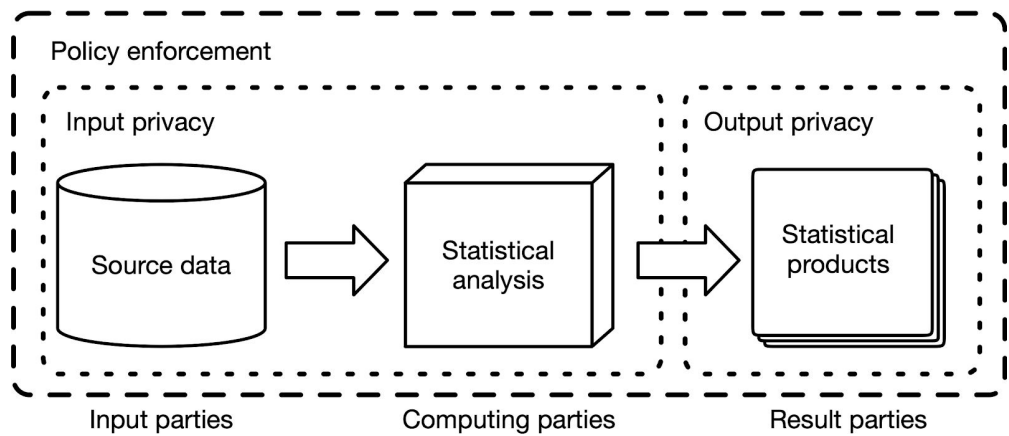

*Figure 4: How multiple privacy goals co-exist in a system*

Input privacy covers source data, the intermediate and final results of processing. Input parties are responsible for protecting their own input data, but once it is transferred, the recipient must continue protecting it.

Output privacy is a property of the statistical products. Even though the computing parties are responsible for ensuring that the results of computation have some form of output privacy, the risks are nearly always related to a result party learning too much.

Policy enforcement covers the whole system – input parties may ask for controls on processing before granting the data, result parties may want to remotely audit the processing for correctness. The responsibility for making such controls available rests with the computing parties who, in our case, are the National Statistics Offices.

# Privacy Enhancing Technologies for Statistics

## Technology Overview

In this handbook, we present multiple Privacy Enhancing Technologies for statistics. For each, we describe the privacy goals they support and how those goals are supported. We consider the following technologies:

1) Secure Multiparty Computation (abbreviated MPC)
2) (Fully) Homomorphic Encryption (abbreviated as HE or FHE)
3) Trusted Execution Environments (abbreviated as TEE)
4) Differential Privacy
5) Zero Knowledge Proofs (abbreviated as ZK Proofs)

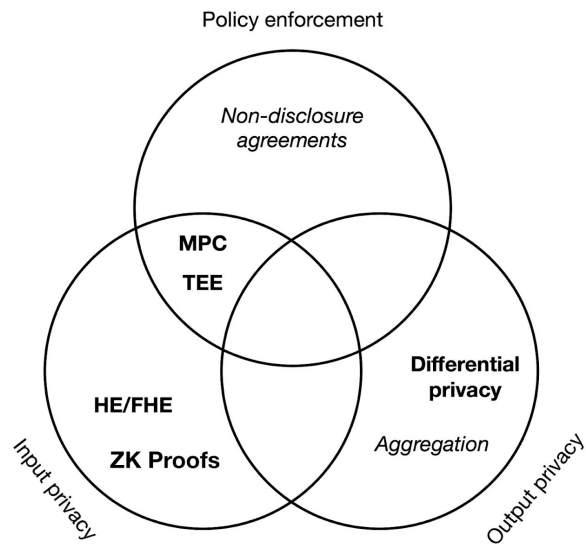

*Figure 5: a Venn diagram showing which privacy goals are fulfilled by which privacy techniques. Techniques in italics are common techniques not included in this handbook.*

Figure 5 shows how the technologies we consider apply to the privacy goals outlined above. The goal of Input Privacy is primarily addressed by secure computation technologies – techniques that compute on data while it remains encrypted or otherwise obfuscated from regular access – and Zero Knowledge proofs of knowledge that prove claims without revealing the input data on which those claims are based. Sometimes these technologies also provide the means to enforce access control policy on a flexible basis.

The goal of Output privacy is primarily addressed by technologies such as differential privacy – techniques that prevent those with access to computed results from "reverse engineering" those results to learn about input data. The goal of Policy Enforcement is primarily addressed by access control policies and enforcement points for those policies – for example by allowing only certain queries over data to be answered by a system. While technology specific to policy enforcement is beyond the scope of this handbook, we note that MPC may enable this capability by enforcing access control *during* secure computation.

Figure 6 shows a top-level Wardley map of the ecosystem of national statistics office (NSO) computation. Wardley maps are widely used to visualise priorities, or to aid organizations in developing business strategy. A Wardley map is often shown as a two-dimensional chart, where the horizontal dimension represents readiness and the vertical dimension represents the degree to which the end user sees or recognizes value. Readiness typically increases from left to right, while recognized value by the end user increases from bottom to top in the chart. Dependencies or hierarchical structure among components is shown by edges among components, each of which is shown as a vertex or point. Wardley maps such as those we use here may be *hierarchical* – that is, a symbol on one Wardley map may reference another "sub-map", allowing representation of more complex dependency structures that might be shown on a single map.

As shown in Figure 6, NSOs are charged to deliver diverse official statistical reports. While these reports often rely on public data, they also sometimes rely on sensitive data from various sources. Use of sensitive data relies on several things shown in the figure. Note that there are no objective completeness criteria for Wardley maps, so Figure 6 may omit certain dependencies in the interest of showing the relationships relevant to this handbook. One dependency for sensitive data is technical access controls that provide the means to keep the data private where necessary. One such area of control is ensuring privacy *during or after computation.* Input privacy and output privacy are key concepts in this area. The technologies we focus on in this handbook fall under these concepts, as shown in the figure.

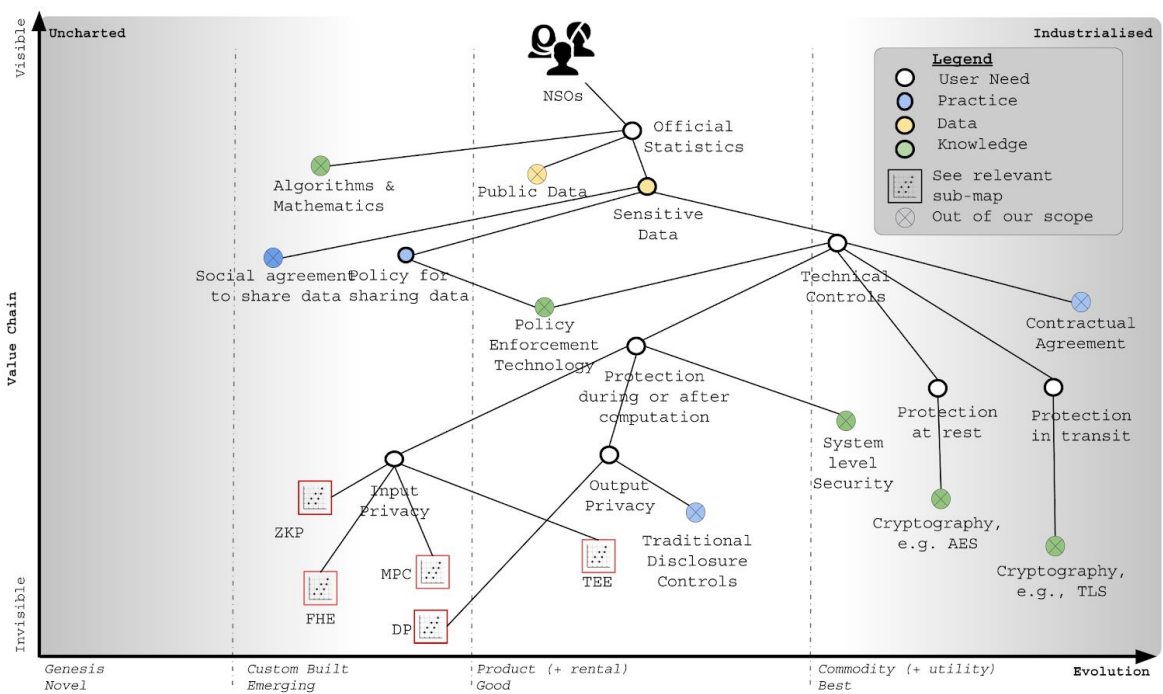

*Figure 6. Top-level Wardley map for privacy-preserving techniques in the context of national statistics offices.*

# Secure Multi-Party Computation

## Overview

Secure multi-party computation (also known as secure computation, multi-party computation/MPC, or privacy-preserving computation) is a subfield of cryptography. MPC deals with the problem of jointly computing an agreed-upon function among a set of possibly mutually distrusting parties, while preventing any participant from learning anything about the inputs provided by other parties[4]; and while (to the extent possible) guaranteeing that the correct output is achieved.

MPC computation is based on secret sharing of computation inputs (and intermediate results). In secret sharing, first introduced by Adi Shamir[5], data is divided into *shares* that are themselves random, but when combined (for example, by addition) recover the original data. MPC relies on dividing each data input item into two or more shares, and distributing these to compute parties. The homomorphic properties of addition and multiplication allow for those parties to compute on the shares to attain *shared results*, which when combined produce the correct output of the computed function. To perform the shared computation required for MPC, all participating compute parties follow a *protocol*: a set of instructions and intercommunications that when followed by those parties implements a distributed computer program.

Modern MPC protocols that tolerate covert or malicious adversaries also rely on zero-knowledge proofs usable by honest players to detect bad behavior (and typically eliminate the dishonest party).

## Examples of Applied Uses

MPC has been applied to many use cases. End-to-end encrypted relational database prototypes use MPC to compute the answers to SQL queries over data that is held only in encrypted form in the database. Statistical analytic languages such as R have been augmented with MPC capability to protect data during statistical and other computations. MPC is used to protect cryptographic key material while using those keys for encryption, decryption, and signing. MPC is also used in streaming data environments, such as processing VoIP data for teleconferencing without requiring any trusted server in the VoIP system. A recent paper describes some of the leading use cases in more detail[6].

---

[4] Other than what can be inferred solely from the function's output.

[5] Adi Shamir. 1979. How to share a secret. Commun. ACM 22, 11 (November 1979), 612-613.

[6] *David W. Archer, Dan Bogdanov, Liina Kamm, Yehuda Lindell, Kurt Nielsen, Jakob Illeborg Pagter, Nigel P. Smart and Rebecca N. Wright.* **From Keys to Databases – Real-World Applications of Secure Multi-Party Computation. https://eprint.iacr.org/2018/450**

One interesting potential application for MPC is for long-term shared data governance. Because MPC relies on cryptographic secret sharing with access control over those shares controlled jointly by all parties involved, data can be stored indefinitely in secret shared form and only recovered if the appropriate proportion of parties agrees. This capability is related to the notion of secret sharing of data at rest, and more distantly related to the notion of *threshold encryption.*

### Adversary Model and Security Argument

Because MPC assumes the possibility of mutually distrusting parties, it also assumes a new class of adversary: one that controls one or more participants in the computation. Such an adversary might be an *insider threat*, or might be a Trojan or other penetrative, long-lived attack from outside an organization. This new class of adversary is typically described in terms of several traits: degree of honesty, degree of mobility, and proportion of compromised compute parties are the typical traits described in the literature.

**Honesty**. In the *semi-honest* adversary model, such control is limited to inspection of all data seen by the corrupted participants, as well as an unlimited knowledge about the computational program they jointly run. In the *covert* model, an adversary may extend that control to modifying or breaking the agreed-upon protocol, usually with the intent of learning more than can be learned from observation alone. However, in this model the adversary is motivated to keep its presence unobserved, limiting the actions it might take. In the *malicious* model, an adversary may also modify or break the agreed-upon protocol, but is not motivated to keep its presence hidden. As a result, a malicious adversary may take a broader range of actions than a covert adversary.

**Mobility.** A *stationary* adversary model assumes that the adversary chooses *a priori* which participants to affect. Such a model might represent for example that one compute participant is compromised, but others are not. Stronger versions of this adversary mobility trait allow for an adversary to move from participant to participant during the computation. At present, a real-world analog of such an adversary is hard to imagine.

**Proportion of compromised parties**. MPC adversary assumptions fall into one of two classes: *honest majority*, and *dishonest majority*.

Just as there are a variety of participant adversary models for MPC, there are also diverse MPC protocols that provide security arguments that protect against those adversaries. Security is typically argued by showing that a real execution of an MPC protocol is indistinguishable from an idealized simulacrum where all compute parties send their private inputs to a trusted broker who computes the agreed-upon function and returns the output. The diverse MPC protocols have different properties that enhance security. Those properties typically described are:

- Input privacy, as already described above
- Output correctness – all parties that receive an output receive a *correct* output
- Fairness – either all parties intended to receive an output do so, or none do
- Guaranteed output – all honest parties are guaranteed to complete the computation correctly, regardless of adversary actions sourced by dishonest parties

While input privacy and output correctness can be guaranteed when a majority of compute parties do not follow the protocol, the combination of all four desirable properties (input privacy, output correctness, fairness, and guaranteed output delivery) can only be guaranteed when the majority of compute parties follow the protocol faithfully.

## History

MPC was first formally introduced as secure two-party computation (2PC) in 1982 (for the so-called Millionaires' Problem), and in more general form in 1986 by Andrew Yao. The area is also referred to as Secure Function Evaluation (SFE). The two-party case was followed by a generalization to the multi-party case by Goldreich, Micali and Widgerson.

It should be noted that MPC uses intercommunication among compute parties frequently. In fact, estimations of run-time for MPC protocols can be quite accurate using communication cost as the only estimating factor (that is, ignoring estimates of computation delay at compute parties entirely). This high reliance on both available network bandwidth and network latency between parties kept MPC mainly a theoretical curiosity until the mid 2000's when major protocol improvements led to the realisation that MPC was not only possible, but could be performed for useful computations on an internet latency scale. MPC can be now considered a practical solution to carefully selected real-life problems (especially ones that require mostly local operations on the shares with not much interactions among the parties). Distributed voting, private bidding and auctions, sharing of signature or decryption functions and private information retrieval are all applications that exhibit these properties [11]. The first large-scale and practical application of multiparty computation (demonstrated on an actual auction problem) took place in Denmark in January 2008 [12].

A characterisation of available commercial and Government MPC solutions would be almost immediately out of date. In addition, cataloguing the plethora of academic MPC research tools would be a futile venture. Instead, we offer here a brief list of some companies that offer "point solutions" that apply MPC. Examples of such systems include the Sharemind statistical analysis system by Cybernetica, and cryptographic key management systems from Sepior and Unbound Tech. Other companies offer design consultancies in specific areas based on MPC technology, for example, Partisia helps design market mechanisms based on MPC on a bespoke basis.

There is also a growing number of public domain complete MPC systems developed by government funded research projects. These are either general libraries, general purpose systems or systems that solve a specific application problem. In each of these three categories, we list the SCAPI library (from Bar-Ilan University), the SCALE-MAMBA MPC-system (from KU Leuven) and the Jana[7] relational database (from Galois Inc.).

## Costs of Using the Technology

MPC technology performance depends heavily on the functions to be securely computed. A typical metric for MPC performance is *computational slowdown* – the ratio of the latency of computation in MPC to the latency of the same computation done without MPC security. For general computation such as the calculations needed to process typical relational database query operators, recent results show a slowdown up to 10,000 times.

While it remains tricky to give guidance on where MPC might be performant and where it might not, we have some general guidelines. Computations that rely heavily on addition, such as summations, are typically faster than general computation, while computations that rely on division or other more complex functions are typically much slower. Computations on integer or fixed-point data are relatively faster than those that rely on floating-point computation. Computations that rely on generative functions such as random number generation are also typically slow.

The table below summarizes real example applications and the typical slowdown seen for those computations.

| Example Provider (National origin) | Description of Key Computation | Key Data Type Used | Typical Computational Slowdown per data element, and asymptotic slowdown behavior |
|---|---|---|---|
| Galois, Inc. (USA) | SQL queries | Integers, fixed-point, strings | Up to 10,000 times, linear scaling with data size |
| Cybernetica (Estonia) | Statistical analysis | Databases of integers, fixed point, floating point, some text support | Up to 10 000 times, linear scaling |

---

[7] Jana uses SCALE-MAMBA as part of its backend.

UN Handbook on Privacy-Preserving Computation Techniques        23

## Availability

For the most part, MPC is still an academic research topic. A few companies use specialized MPC protocols for specific functions, and a few specialize in standard or bespoke product development or solutions consulting in this technology. Offerings are constantly advancing, so providing a catalog in this paper would be of limited value.

## Wardley Map for MPC

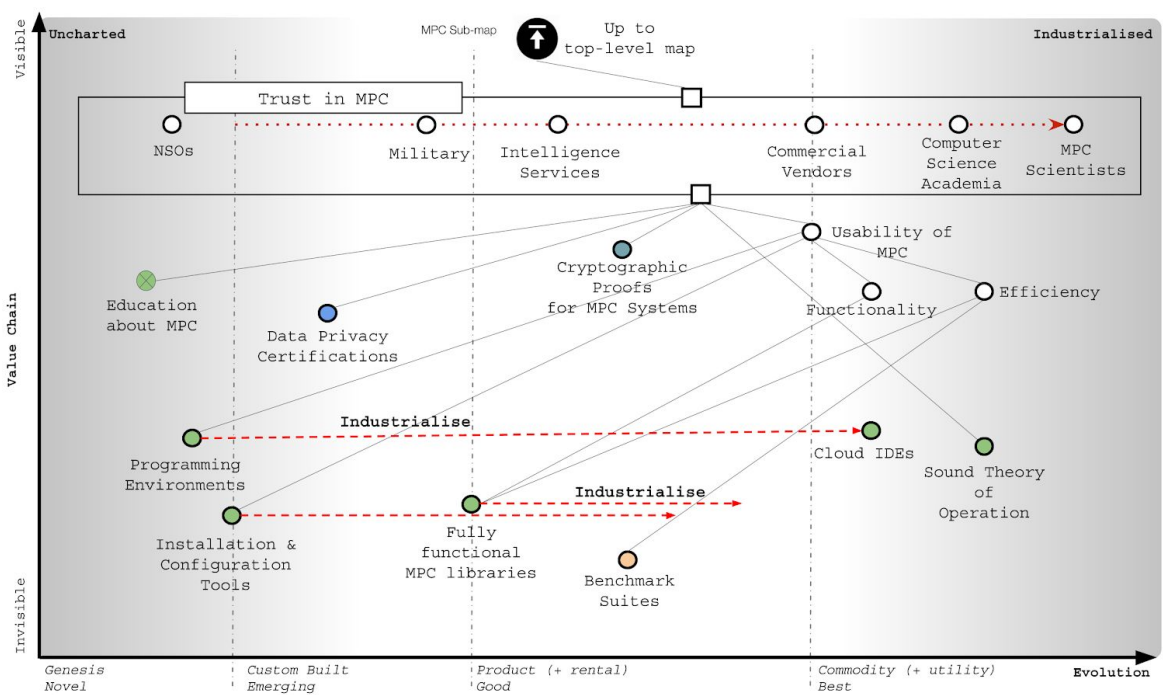

*Figure 7. Wardley map for Multi-party computation.*

Figure 7 presents a Wardley map focused on the details of multi-party computation. While the theory of operation for MPC is at a relatively high state of technology readiness, most of what an end user expects of a computing product is still very early in development. Ease of programming is highly visible to end users of a programming system such as MPC, yet not much has been done to develop the required capabilities. Similarly, MPC is difficult to configure correctly at present, and currently requires highly customised client software as well as server software for deployment. While proof-of-concept demonstrators have shown that these important capabilities *can* be developed, development in a product sense is at a very early stage. Similarly, the assurance story that should give confidence to adopters that MPC technology works correctly without fail is very early in development.

Somewhat further along in readiness is the ability to scale MPC to practical computations. However, many aspects of such computations are still idealised. For example, MPC can execute queries over a relational database, but only for a limited subset of relational queries and data types, and MPC is unable to accommodate important related operations such as data cleaning.

Performance of MPC systems against simplified computations is somewhat further developed, with fieldable prototypes for carefully chosen applications. However, performance remains a challenge, with slowdown factors of 100X up to 100,000X or more compared to "in the clear" computation.

We need to see improvements in MPC education and privacy certifications to improve NSOs trust in MPC products and services.

# Homomorphic Encryption

## Overview

Homomorphic encryption refers to a family of encryption schemes with a special algebraic structure that allows computations to be performed directly on encrypted data without requiring a decryption key. Encryption schemes that support one single type of arithmetic operation (addition or multiplication) have been known since the 1970's[8] and are often said to be *singly* or *partially* homomorphic. The practical value of such a "homomorphic property" was immediately recognised and explored by Rivest, Adleman, and Dertouzos.[9] In 2009 Craig Gentry described the first so-called *fully* homomorphic encryption scheme[10] that allows both additions and multiplications to be performed on encrypted data. This was a significant invention, because in principle such an encryption scheme can allow arbitrary Boolean and arithmetic circuits to be computed on encrypted data without revealing the input data or the result to the party that performs the computation. Instead, the result would be decryptable only by a specific party that has access to the secret key – typically the owner of the input data. This functionality makes homomorphic encryption a powerful tool for cryptographically secure cloud storage and computation services and also a building block for higher-level cryptographic primitives and protocols that rely on such functionality.

---

[8] Ronald Rivest, Adi Shamir, and Leonard Adleman. **A method for obtaining digital signatures and public-key cryptosystems**. *Communications of the ACM*, *21*(2) (1978): 120-126.

[9] Ronald Rivest, Leonard Adleman, and Michael L. Dertouzos. **On data banks and privacy homomorphisms**. *Foundations of secure computation* 4.11 (1978): 169-180.

[10] Craig Gentry and Dan Boneh. **A fully homomorphic encryption scheme**. Vol. 20. No. 09. Stanford: Stanford University, 2009.

While theoretically powerful and academically interesting, the first homomorphic encryption schemes quickly turned out to be unusable in terms of performance and key size. A significant amount of work was done over the next few years in inventing and implementing both simpler and faster homomorphic encryption schemes. This work culminated in the release of the homomorphic encryption library *HElib*[11] by IBM Research, which improved the performance over prior homomorphic encryption implementations by several orders of magnitude. Today there are multiple open source homomorphic encryption libraries available implementing a variety of homomorphic encryption schemes suitable for different applications.

## Note About Terminology

While in principle fully homomorphic encryption schemes allow arbitrary computation on encrypted data, in practice almost all efficient implementations use a so-called *levelled mode* where the encryption scheme is configured to support computations of only a specific or bounded size, typically resulting in significant performance improvements. For simplicity, in this handbook we freely use the term Homomorphic Encryption (HE) to refer to either Fully Homomorphic Encryption (FHE) or Levelled Fully Homomorphic Encryption.

## Examples of Applied Uses

Homomorphic encryption offers powerful post-quantum secure encryption and unique non-interactive encrypted computation functionality, but can result in a high computational overhead and large message expansion. Thus, ideal applications have a relatively small but critical encrypted computation component, include a persistent storage aspect, and are hard or impossible to implement using other methods.

A commonly cited class of applications is in the medical domain, where regulations enforce strict patient data privacy measures, but hospitals and medical clinics may nevertheless want to enable third-party service providers to analyze, evaluate, or compute on their data without engaging in costly and time-consuming legal processes. For example, a service provider may offer an image analysis service for detecting tumors in MRI scans. A predictive model can be evaluated directly on homomorphically encrypted data, avoiding the issue of medical data leaking to the service provider.

For data storage providers a potential application is in performing analytics on encrypted customer data. For example, a customer may want to store a large encrypted database using a cloud storage service and not have to download the entire database for simple computational queries, as this creates unnecessary logistical challenges and potentially exposes the full dataset to a potentially low security computation environment. Instead, all possible aggregation of the data should be performed in encrypted form directly by the cloud storage provider avoiding unnecessary exposure of the data to the client's machine.

---

[11] Shai Halevi and Victor Shoup. **Design and implementation of a homomorphic-encryption library**. IBM Research (Manuscript) 6 (2013): 12-15.

Another promising application is in private set intersection and private information retrieval protocols. In private set intersection a client and a server hold sets of unique identifiers (e.g., names, email addresses, phone numbers) and wish to find the common items in their sets. For example, two companies may want to find the customers they have in common. Homomorphic encryption can yield an efficient solution to this problem when one of the sets is much smaller than the other one. In this case the smaller set can be encrypted homomorphically and sent to the other party, who can evaluate a circuit for matching the encrypted items to its set. The encrypted result remains proportional in size to the smaller set and can be sent back for decryption and interpretation. In private information retrieval one of the parties can additionally retrieve data associated to the matching items without the data owner learning what data, if any, was retrieved. In protocols of this type the data and set sizes must always be upper bounded by publicly known upper bounds, and all communication and computation must be proportional to these upper bounds.

## Adversary Model and Security Argument

Today, all homomorphic encryption schemes with practical – or close to practical – performance are based on the *Learning With Errors*[12] (LWE), or *Ring Learning With Errors*[13] (RLWE) problems. In other words, one can show that if these homomorphic encryption primitives can be efficiently broken, then either LWE or RLWE can be efficiently solved for specific parameterisations. As LWE and RLWE have been studied extensively and are believed to be infeasible to solve by modern computers for these parameterisations, there is strong reason to believe that the corresponding homomorphic encryption schemes are secure.

As homomorphic encryption refers only to a type of encryption primitive and not a protocol, its security definition states merely that, given a ciphertext, an adversary without the secret key cannot obtain any information about the underlying plaintext. This property holds even if the adversary is allowed to obtain any number of encryptions of plaintexts of its own choosing. However, it does not hold when the adversary is allowed to obtain any information about a decryption of a data payload of its own choosing. Indeed, for secure uses of homomorphic encryption it is critical that no information about decrypted data is ever communicated back to the source of the corresponding encrypted data, unless that source is trusted not to misbehave; this includes seemingly innocuous information, such as a request to repeat a protocol execution, refusing to pay for a service, or revealing any change in behavior that can be expected to depend on the outcome of the encrypted computation. The presence of such a back communication channel can at worst result in a full key recovery attack, and at best in a lowered security level. Therefore, outsourced storage and computation involving a single data owner should be considered as the primary use-case of homomorphic encryption. After receiving the result, the secret key owner must not perform any action that is observable to the service provider based on the decrypted result to avoid the attacks described above.

---

[12] Oded Regev. **On lattices, learning with errors, random linear codes, and cryptography**. *Journal of the ACM (JACM)* 56.6 (2009): 34.

[13] Vadim Lyubashevsky, Chris Peikert, and Oded Regev. **On ideal lattices and learning with errors over rings**. *Journal of the ACM (JACM)* 60.6 (2013): 43.

Another subtlety is that most homomorphic encryption schemes do not provide *input privacy*: if a computation depends on the private encrypted input of two or more parties, the encryption scheme is not guaranteed to protect these inputs *from the owner of the secret key*. Homomorphic encryption is also *malleable* by nature, so anyone intercepting a ciphertext can be expected to modify the underlying plaintext unless, e.g., the ciphertext is cryptographically signed by the sender.

It is important to understand that homomorphic encryption is a low level cryptographic primitive and building secure protocols from it is not possible without the help of a cryptography expert. Even in the simplest cases such protocols can result in unexpected or unintended security gaps. Most homomorphic encryption based protocols can be proved to be secure only in a semi-honest security model, although there are exceptions where a stronger security model is achieved by combining homomorphic encryption with other primitives.[14]

## Costs of Using the Technology

The use of homomorphic encryption comes with at least three types of costs: message expansion, computational cost, and engineering cost.

In HE systems, encrypted data is typically significantly larger than unencrypted data due to encoding inefficiency (converting real data into plaintext elements that can be encrypted) and inherent expansion from the encryption scheme (ratio of ciphertext size to plaintext size). Depending on the use-case, encoding inefficiency can range from the ideal case (no expansion at all) to an expansion rate measured in the tens or hundreds of thousands when the encoding method is poorly chosen. The inherent message expansion can in principle be arbitrarily large, but in practice expansion factors of 1–20x can be expected, depending on the use case. Thus in most cases, one should not think of encrypting large amounts of data with homomorphic encryption, but instead carefully consider what data exactly will be needed for the desired encrypted computations and encrypt only that.

---

[14] Hao Chen, Zhicong Huang, Kim Laine, and Peter Rindal. **Labeled PSI from Fully Homomorphic Encryption with Malicious Security**. *Proceedings of the 2018 ACM SIGSAC Conference on Computer and Communications Security*. ACM, 2018.

The computational cost of homomorphic encryption is significant *compared to unencrypted computation*. The exact cost depends strongly on the parameterisation of the encryption scheme and whether throughput or latency is measured. Namely, most homomorphic encryption schemes support natively high-dimensional vectorized computations on encrypted data, and if this vectorisation can be fully utilised it can increase the throughput by 1,000–100,000x. On a CPU, additions of plaintext elements (i.e., ignoring the improvement from inherent vectorisation) can be expected to be performed in 10–10,000μs and multiplications can be expected to be 100–500x slower.

Developing complex systems with homomorphic encryption can be challenging and should always be done with the help of an expert, making the initial cost for such solutions potentially high. There are two reasons for this: the security model – as discussed earlier – can be hard to comprehend and evaluate without special expertise, and the available homomorphic encryption libraries can be hard to use to their full potential without deep understanding of how they work. It should also be noted that homomorphic encryption can be hard or impossible to integrate with existing systems. Instead, sophisticated applications of this technology can require substantial changes in existing data pipelines, data manipulation procedures and algorithms, and data access policies. Of course, this is equally true for most security/privacy techniques and will not come as a surprise to advocates of modern security-conscious development strategies such as the *rugged DevOps*, where security/privacy considerations are inherently built into a continuous integration and continuous deployment strategy.

## Availability

The most commonly used (fully) homomorphic encryption schemes at this time are the Brakerski-Gentry-Vaikuntanathan (BGV) and the Brakerski-Fan-Vercauteren (BFV) schemes. Both allow encrypted computation on vectors of finite field elements. More recently the Cheon-Kim-Kim-Song (CKKS) scheme has gained popularity but is still relatively little known partly due to small number of implementations. The CKKS scheme allows approximate encrypted computation on real or complex numbers, which can be well suited for statistical and machine learning applications. The trade-offs between the different schemes are complicated and can be difficult to understand even for experts in the field. For very large and very small computations the BGV scheme has a performance advantage over the BFV scheme, but in many other cases the difference is negligible with modern optimisation techniques. On the other hand, the BGV scheme is more complicated and has a steeper learning curve than the BFV scheme. The CKKS scheme is comparable in performance to BGV but can be even more challenging to learn. However, it provides functionality that is not available from other schemes.

The BGV scheme is implemented in the HElib library from IBM Research and in the PALISADE library/framework from the New Jersey Institute of Technology. BFV is implemented in the Microsoft SEAL, PALISADE, and FV-NFLlib libraries. CKKS is implemented in Microsoft SEAL, HEAAN, and HElib.

While BGV, BFV, and CKKS all allow in theory arbitrary computation on encrypted data, they are typically far more efficient to use in a leveled mode where the depth of the circuit is predetermined and the parameters of the encryption scheme are chosen to enable computations up to precisely this depth. The Torus FHE (TFHE) scheme instead operates on bit-wise encrypted inputs and attempts to optimise for enabling arbitrary computation. A scheme such as TFHE can be the most efficient solution in cases where bitwise encrypted input is necessary, such as in computations involving comparisons of encrypted numbers, sorting, or similar non-polynomial operations. The TFHE scheme is implemented a library with the same name.

## Wardley Map

Figure 8 presents a Wardley map focused on the details of homomorphic encryption (HE).

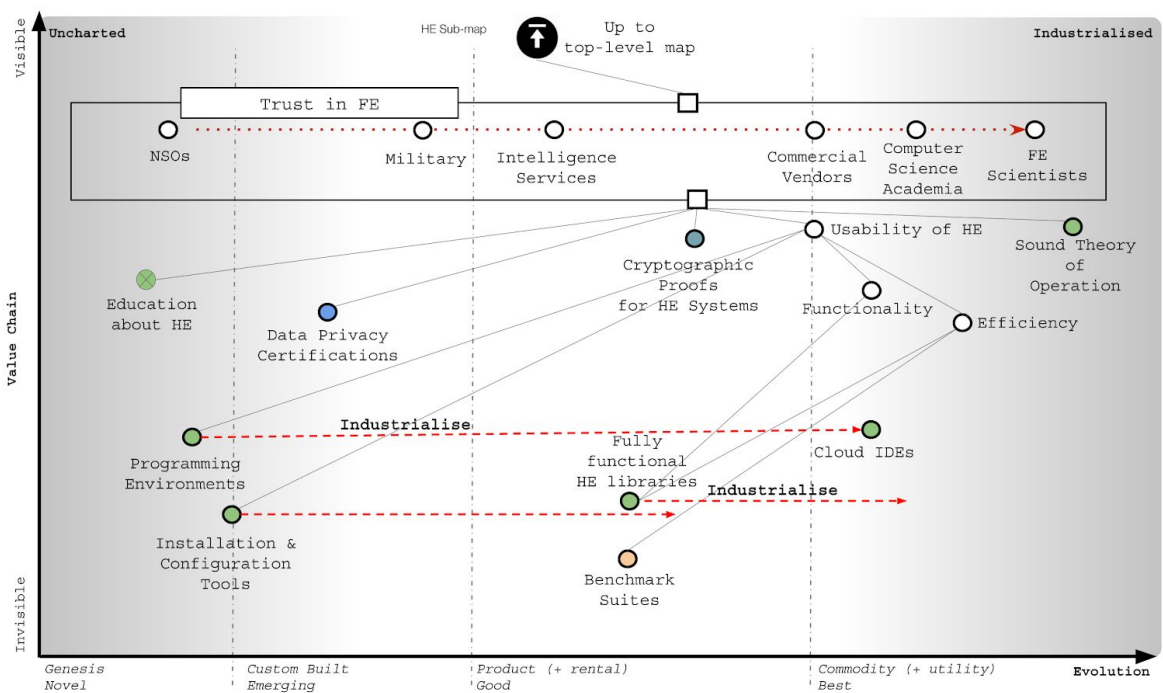

*Figure 8. Wardley map for Homomorphic Encryption Technologies.*

While the theory of operation for HE is at a high state of technology readiness, applicable technology solutions are still rare. Ease of programming is still at the level of a few libraries with increasingly useful programmer interfaces, yet no strong development environments exist outside academic research. We point to the RAMPARTS system, developed by Galois, Inc. and the New Jersey Institute of Technology for IARPA in the USA, as one example of what such a programming environment might look like for general purpose programs in a scientific analytic programming language. Similarly, HE is difficult to configure correctly, because the correct choice of security parameters depends on the complexity of the computation to be performed.

Another current concern with HE solutions is the lack of verifiability of computation. Because outsourcing of computation is expected to be a common use model for HE, it may be important for the client of the computation to be able to verify that the computation was executed correctly. This capability is not available in HE solutions today.

Somewhat further along in readiness is the ability to scale HE to practical computations. While slowdown factor of individual computations is still in the hundreds of thousands, parallelism is a natural outcome of HE, potentially enabling slowdowns to improve into the hundreds in cases where thousands of computation instances are wanted, such as in the SIMD (single instruction, multiple data) execution paradigm. In addition, significant work has been done to leverage hardware accelerators such as GPGPUs (General-purpose Graphics Processing Units) to accelerate such parallel computations.

We need to see improvements in FE education and privacy certifications to improve NSOs trust in FE products and services.

## Differential Privacy

### Overview

Differential privacy (DP) provides an information-theoretic notion of Output Privacy. Its goal is to quantify and limit the amount of information about individual records in a database that is leaked by releasing the result of an aggregate computation on that database. DP was first introduced in 2006 by Dwork et al.[15]. Historically, DP is related to the privacy models classically studied in the literature on statistical disclosure control and statistical databases. DP provides a more general notion of privacy than other specialized definitions such as k-anonymity, which focuses on the context of data anonymization. Furthermore, DP was designed to avoid pitfalls that previous attempts to define privacy suffered, especially in the context of multiple releases and when adversaries have access to side knowledge. We note that such pitfalls also affect less sophisticated attempts at privacy preservation, such as aggregation alone and *ad hoc* noise addition to aggregate results.

---

[15] Dwork, Cynthia, Frank McSherry, Kobbi Nissim, and Adam Smith. **Calibrating noise to sensitivity in private data analysis.** In *Theory of cryptography conference*, pp. 265-284. Springer, Berlin, Heidelberg, 2006.

Differential privacy specifies a property that a data analysis algorithm must satisfy in order to protect the privacy of its inputs. In this sense, DP is a privacy standard, rather than a single tool or algorithm. The DP property is stated in terms of an alternate world where the input of a particular individual has been removed from the database. DP requires that the outputs produced by the algorithm in the real and alternate world are statistically indistinguishable. Being a property of the algorithm means that such indistinguishability must hold regardless of what the database is and which individual we choose to remove. DP is therefore not a property of the output, and cannot be measured directly by looking at the output of the algorithm on a given input database. Another crucial remark about the definition of DP is that the indistinguishability requirement is too strong to be satisfied by any deterministic algorithm. Randomness is therefore an indispensable ingredient in the design of any differentially private algorithm.

The versatility and robustness of the principles behind differential privacy has led to a number of variations of the basic definition, in addition to a few different threat models. The two most important threat models lead to what is usually called local DP and curator DP. In the local model, differential privacy is applied directly to the data by each individual before it is collected and aggregated. In the curator model, a trusted party collects data from a number of individuals and then runs a differentially private data analysis algorithm whose output is released. We note that the curator model can be combined with input privacy-preserving techniques such as multi-party computation, where the MPC technique protects the input data.

The interested reader should consult the recent paper by Nissim et al.[16] for a more extensive non-technical introduction to DP. Additionally, monographs by Dwork and Roth[17] and Vadhan[18] provide a comprehensive account of the basics of differential privacy from a technical perspective.

## Examples of Applied Uses

Differential privacy is just over 12 years old as of this report, making it a relatively new development in the field of privacy-enhancing technologies. The last decade has seen a fast growing interest on research in theoretical and algorithmic aspects of DP. Some generic DP systems in the context of database interfaces and synthetic data release methods have been proposed in the literature, but virtually none of these systems come with production-ready implementations. However, the interest generated by the solid principles behind DP and the growing concerns about online privacy have led to a small number of real-world deployments, typically using ad-hoc algorithms for specific applications.

---

[16] Kobbi Nissim, Thomas Steinke, Alexandra Wood, Micah Altman, Aaron Bembenek, Mark Bun, Marco Gaboardi, David O'Brien, and Salil Vadhan. **Differential Privacy: A Primer for a Non-technical Audience.** Vanderbilt Journal of Entertainment and Technology Law. Forthcoming.

[17] Dwork, Cynthia, and Aaron Roth. **The algorithmic foundations of differential privacy**. Foundations and Trends® in Theoretical Computer Science 9, no. 3–4 (2014): 211-407.

[18] Vadhan, Salil. **The complexity of differential privacy.** In Tutorials on the Foundations of Cryptography, pp. 347-450. Springer, Cham, 2017.

Two well-known applications of DP are its use in Google Chrome and Apple's iOS/OSX to collect usage statistics in a privacy-preserving way. These applications follow the local model of DP, where each individual user privatizes their own data before sending it to a centralized server for analysis. For example, Chrome used this approach to discover frequently visited pages in order to improve its caching and pre-fetch features, while iOS uses it to discover words and emojis frequently used in a texting application to improve the language models used in typing assistance. Additionally, Microsoft also announced that they employ DP in the local model to collect telemetry data from devices running their operating systems.

In the curator model of differential privacy, the most well-known usage is by the U.S. Census Bureau, who are planning to use it when releasing the results of the upcoming 2020 Census. This was motivated by research showing that without the kind of protection provided by differential privacy it is sometimes possible to recover accurate information about individuals from Census data even though only aggregates at different levels of granularity are released.

## Adversary Model and Security Argument

Differential privacy offers a mathematical guarantee to individuals contributing sensitive data to a database on which certain queries will be performed. The guarantee takes the form of a bound on the risk incurred by individuals contributing their data, and builds upon the intuition that queries that are invariant to removal of any single record of a database are immune to membership and reconstruction attacks (regardless of the side-knowledge of the adversary). In other words, differential privacy provides a convincing argument for a user to contribute data to a database, as it guarantees that query results will be very similar regardless of whether the user joins the database or not. This quality protects against attacks where an adversary is allowed to query the database and has access to unlimited side knowledge. The fact that DP withstands such powerful attackers can be counterintuitive, but in fact emphasizes the fact that differential privacy quantifies the leakage of an algorithm, and it is not a property of the data.

Technically, DP is formalized by saying that a mechanism for releasing the result of a query on a database is differentially private if an adversary observing this release will not be able to determine if any particular record was present in the database. This guarantee takes a statistical flavor: since DP requires that the data analysis algorithm must be randomized, the adversary's inability of determining the presence of a record in the database is measured in terms of the similarity between the probability distributions over outputs when the record is either present or missing in the database. This similarity measure is parametrized numerically (typically represented by greek letters epsilon and delta), with smaller values for the parameters representing a stronger privacy protection. Although these values have a very precise statistical interpretation, there is no general application-agnostic recipe for choosing appropriate values of these parameters -- one of the current limitations in usability of DP.

As explained above, DP provides privacy even in the context of adversaries with access to arbitrary side-knowledge. Additionally, no assumptions are required either on the computational capabilities of an adversary. The exact threat model under which DP is used depends on the trust assumptions imposed on the system. This leads to the local and curator models described above, which differ in that the latter assumes a trusted party will collect the data to be analyzed and release the results of such analysis using differential privacy, while the former makes no trust assumptions on the entity collecting the data.

## Costs of Using the Technology

The main cost of using differential privacy is a loss in terms of accuracy with respect to solutions for the same problem that do not provide output privacy. Typically, this cost depends on the level of privacy required (more privacy incurs more loss in accuracy), the amount of data available (increasing the amount of data available reduces the accuracy loss), and the threat model (for the same problem, local DP generally loses more accuracy than curator DP).

In addition, the cost of differential privacy is also influenced by the amount of information being released. For example, releasing few statistics about a dataset can typically be done with more accuracy than releasing a more complex object, such as a machine learning model or a synthetic dataset. Moreover, an important observation that motivates the definition of differential privacy is that one can't hope to query a database indefinitely without ultimately revealing a large percentage of its contents. This makes deployments where the queries are not fixed *a priori* especially challenging for DP applications.

On the computational side, differential privacy generally incurs only a moderate increase in complexity over non-private alternatives. However, there are also important problems where DP algorithms leading to the best-known accuracy suffer from scalability issues or are intractable.

## Availability

As of this writing, there are no freely or commercially available products that are generic and production-ready. The following contain pointers to academic resources that implement DP for a variety of problems, including training of machine learning models and querying relational databases.

- https://www.microsoft.com/en-us/research/project/privacy-integrated-queries-pinq/
- https://github.com/uber/sql-differential-privacy

- https://github.com/tensorflow/privacy
- http://www.bipr.net/diffpriv/

The Private data Sharing Interface (PSI) developed by the Privacy Tools project led by Harvard University implements a generic methodology for providing differentially private access to sensitive datasets. PSI focuses on typical use cases in the social sciences, allowing researchers to upload sensitive datasets, release a set of selected statistics with differential privacy, and allow other researchers to create their own DP queries against the dataset. The tool comes with a graphical user interface that guides data owners through the release process, helping them create an appropriate privacy budget and also select from a number of readily available statistics.

### Wardley Map

Figure 9 presents a Wardley map focused on details of Differential Privacy.

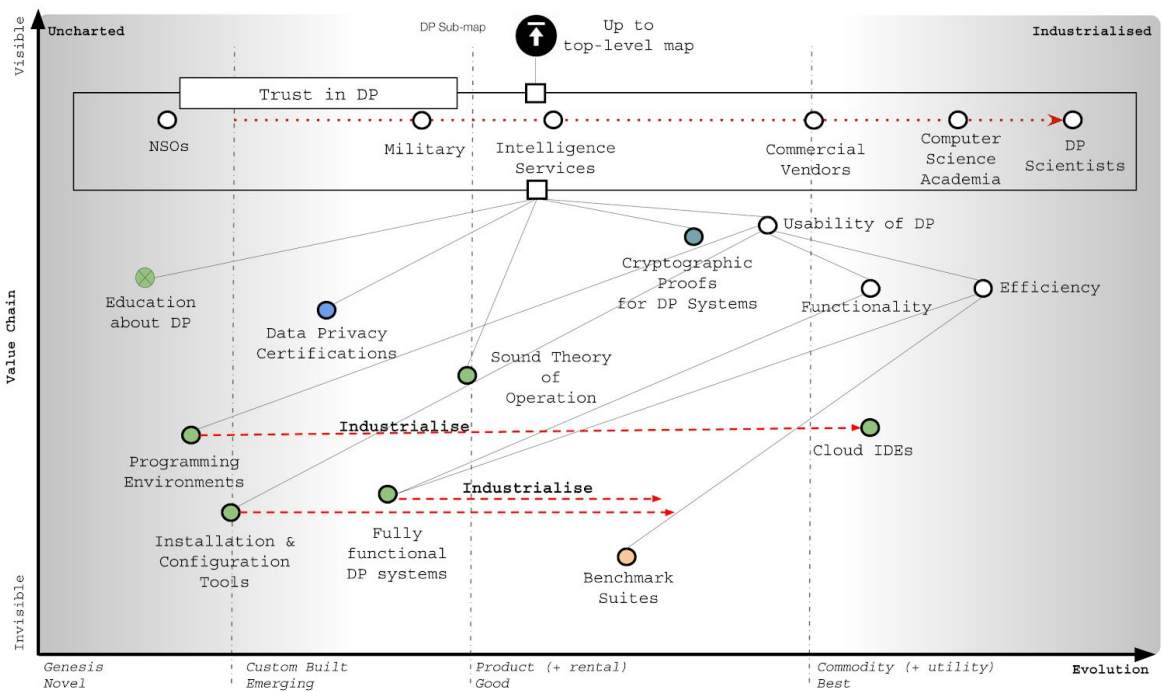

*Figure 9. Wardley map for Differential Privacy Technologies.*

Differential privacy is relatively hard to characterise in terms of readiness for use. DP focuses on output privacy: protection that prevents input data from being learned by parties observing the output. Thus traditional cryptographic security arguments do not apply to DP because information leakage is a function of what questions are asked of the data and how often they are asked. As a result, DP may be quite secure and well understood in environments where the data owner is the only querier and releases only DP-protected aggregate results. However, in settings where other users are able to pose queries, the data owner must establish privacy budgets -- a part of the DP research ecosystem that is at present poorly understood.

Because of the above complexity in characterizing use settings for DP, education in how to correctly use DP systems is lacking. In addition, few production systems exist outside academia, and those that we know of rely on the "owner runs the queries" model described above for security.

We need to see improvements in DP education and privacy certifications to improve NSOs trust in DP products and services.

# Zero Knowledge Proofs

## Overview

Zero knowledge (ZK) proofs are a cryptographic technology that allows one party (called the *prover*) to prove statements to another party (called the *verifier*) that depend on secret information known to the prover without revealing those secrets to the verifier. A simple example of such a statement is "I am an adult at least 21 years old". A more complex statement might require running a machine learning prediction model on the whole portfolio and past transaction history of a company to prove its solvency, and do so without revealing any of that sensitive data.

A zero-knowledge proof has three salient properties:

- **Completeness:** If the statement is true and both the prover and the verifier follow the protocol; the verifier will accept the proof.

- **Soundness:** If the statement is false, and the verifier follows the protocol; the verifier will not be convinced by the proof.

- **Zero-knowledge:** If the statement is true and the prover follows the protocol; the verifier will not learn any confidential information from the interaction with the prover except that the statement is true.

Zero knowledge proofs (also called *zero knowledge arguments)* were introduced in the work of Goldwasser, Micali and Rockoff[19]. While there is a technical difference between these two terms

---

[19] **The knowledge complexity of interactive proofs**, Shafi Goldwasser, Silvio Micali, Charlie Rackoff, SIAM Journal of Computing, Vol. 18, 1989

(in terms of whether their security guarantees hold against computationally bounded or unbounded adversaries), we use the notions here interchangeably.

There are different types of zero knowledge constructions in terms of setup requirements, efficiency and the interactiveness of algorithms, proof succinctness and the hardness assumptions required.

- **Type of statements supported:** some ZK proofs support arbitrary statements, i.e. the prover can prove any computation on its secret input. Other proofs are tailored for very specific statement, e.g. knowledge of a specific discrete logarithm of a secret.

- **Trusted setup:** some ZK systems require a setup phase. This setup must be *trustworthy*, done either by a trusted party or by running a secure computation between the participants. For example, in the context of the Zcash cryptocurrency, this was done in the *Powers ot Tau Ceremony* https://www.zfnd.org/blog/conclusion-of-powers-of-tau/).

- **Interactiveness:** Some zero-knowledge systems require that the prover and verifier interact during the verification of the proof. Others enable the prover to generate locally a complete proof, which is sent to the verifier who verifies it locally.

- **Efficiency:** There are different measurements for efficiency in a zero-knowledge protocol. These include the length of the proof and the computation complexity of the prover and the verifier. Non-interactive zero knowledge systems that have proofs that are constant in size and a verifier effort that is roughly constant. Such proof systems are called succinct non-interactive zero-knowledge proofs (SNARKs). Such systems usually require additional overhead on the part of the prover. Systems that are interactive, have longer proofs or require more work on the verifier's side usually achieve lower overhead for the prover.

- **Assumptions:** SNARKs require a type of hardness assumption called *non-falsifiable* assumptions[20]. Such assumptions have been accepted and are used in many systems. However, other zero knowledge systems that do not achieve the same efficiency as SNARKs rely on more standard cryptographic hardness assumptions.

## Examples of Applied Uses

In recent years there has been an increasing number of practical applications that leverage zero knowledge proofs. A lot of these applications have been motivated in the context of blockchains where zero knowledge provides the capability to add encrypted transactions to the ledger and then prove that those are consistent with the available resources of the parties or are compliant with regulations governing these exchanges. Effectively zero knowledge brings privacy to public

---

[20] **Separating succinct non-interactive arguments from all falsifiable assumptions**, Craig Gentry, Daniel Wichs, STOC, 2011

ledgers while preserving all desirable verifiability properties. The crypto currency ZeroCash [21] was one of the first adopters of this functionality. Currently there are numerous companies that offer product in this space including Difinity, QED-it, R3, and others.

Zero knowledge proofs provide auditing mechanisms in settings where the underlying information is private and it should not be revealed in full to the auditor. These techniques have potential applications in various contexts:

- checking that taxes have been properly paid by some company or person;
- checking that a given loan is not too risky;
- checking that data is retained by some record keeper (without revealing or transmitting the data);
- checking that an airplane has been properly maintained and is fit to fly.

In many of the above auditing and compliance checking scenarios the underlying computation is a data analysis algorithm. Thus zero knowledge enables proofs that a given output is the output of a correct data analysis on some sensitive input data.

## Adversary Model and Security Argument

Zero knowledge proofs provide two types of guarantees: on the one hand, the successful verification of the proof guarantees that the statement that the prover claims must hold, i.e., no prover can generate a cheating proof. On the other hand, the proof does not reveal any further information about the private input of the prover beyond what the statement reveal (i.e., a zero knowledge proof that a committed amount is above 100, reveal nothing more about the exact number). However, if the verifier has some input for the computation of the statement, the prover does learn this input (i.e., if the prover has a proprietary genetic testing algorithms and the verifier wants to learn the evaluation of the this algorithm on his DNA information together with a proof for the validity of the output, the prover will need to learn the DNA input). Providing privacy for the verifier's input is a more challenging problem. There are ways to achieve this combining zero knowledge proofs with secure computation or fully homomorphic encryption techniques, but it comes at a substantial efficiency cost. We do not address that topic further in this handbook.

The security properties of zero knowledge proofs are based on mathematical hardness assumptions. Diverse ZK systems rely on different assumptions. One notable example is the succinct non-interactive proof, which relies on *non-falsifiable* assumptions -- a special type of assumptions where we cannot efficiently verify if an adversary has broken the assumption. While not typically considered a class of standard cryptographic hardness assumptions, such assumptions are used by many ZK systems.

---

[21] ZeroCash http://zerocash-project.org/

The appeal of succinct non-interactive arguments (SNARGs) comes from efficiency guarantees for the proof length and the verifier and the non-interactive verification algorithm, which are often crucial requirements in systems where the prover and the verifier cannot be online at the same time, and efficient verification is the bottleneck for the system's efficiency. These efficiencies come at the price of non-falsifiable assumptions and often increased prover's complexity. Most interactive proof system can be converted into non-interactive using the so-called "Fiat-Shamir" heuristic which assumes from hash functions with ideal properties, which are known as random oracles. This is just a heuristic since we cannot achieve these ideal properties from any concrete hardness assumption. Thus, achieving non-interactiveness in this manner also has implications for the strength of the security argument. The "Fiat-Shamir" heuristic has also been widely used for practical applications.

Most zero knowledge systems are proven in the setting of a single execution where at any time the prover is executing a single proof with a single verifier, and similarly the verifier is interacting with a single prover. Such security proof does not guarantee that the system preserves its security properties in concurrent executions, where there are many proofs being executed in parallel. Such concurrency issues are mostly relevant for interactive ZK proofs.

## Costs of Using the Technology

There are several costs to consider when using a zero knowledge system. These include efficiency of the proof generation by the prover, efficiency of proof verification by the verifier, size of the proof, and whether the verification requires interaction between the prover and the verifier. For example, SNARG(SNARK) systems provide proofs of small constant size (usually a few hundred bytes), which requires very little communication between prover and verifier. Verification is very efficient, usually taking a few milliseconds (dependent on the length of the input from the verifier). However, the SNARK prover incurs overhead for the computation of the proof. Usually the runtime for this computation is several orders of magnitude slower than the computation of the statement "in the clear".

There are also many other types of zero knowledge systems apart from SNARGs. They offer different trade-offs: they may require interaction between the prover and the verifier, may have longer proofs (logarithmic, square root or linear in the statement size), or be more expensive to verify. The main advantage of such systems is that they impose substantially less overhead on the prover, which is useful in cases when this is the bottleneck for the application.

There are also zero-knowledge systems that are specialized for only particular types of statements, for example, knowledge of credential associated with a public commitment for an identity, which is used in the context of anonymous credentials. These systems might be more efficient in certain applications than using directly general zero knowledge systems.

## Availability

Most of the existing freely accessible implementations of zero knowledge proof systems are implementations accompanying academic papers. Most of these systems can only prove relatively simple statements such as matrix multiplication, hashing, verification for a Merkle tree, or the correctness of a machine learning model inference. Users should be experts in the field and should be aware of the subtleties and the differences of the guarantees the systems provide. The following link provides proiter to the main zero knowledge construction and their corresponding implementations:

https://zkp.science/

In recent years there has been a strong push for adoption of zero knowledge in real software applications. Over the years, a number of companies have built products relying on ZK capabilities, including such UProve (from Microsoft) and Idemix (from IBM). The first main practical application for zero knowledge has been in the context of cryptocurrencies such as zCash and more broadly in blockchains. There is also an initiative for standardization of zero knowledge techniques and constructions. The first zero knowledge standardization workshop was held in May 2018 and the second one will be in April 2019. The following link provides information about this effort as well as proceedings from the workshops, it also lists industrial participant who in most cases have product related to this technology. We note that in the USA, NIST recently held a meeting on ZK standardization.

## Wardley Map

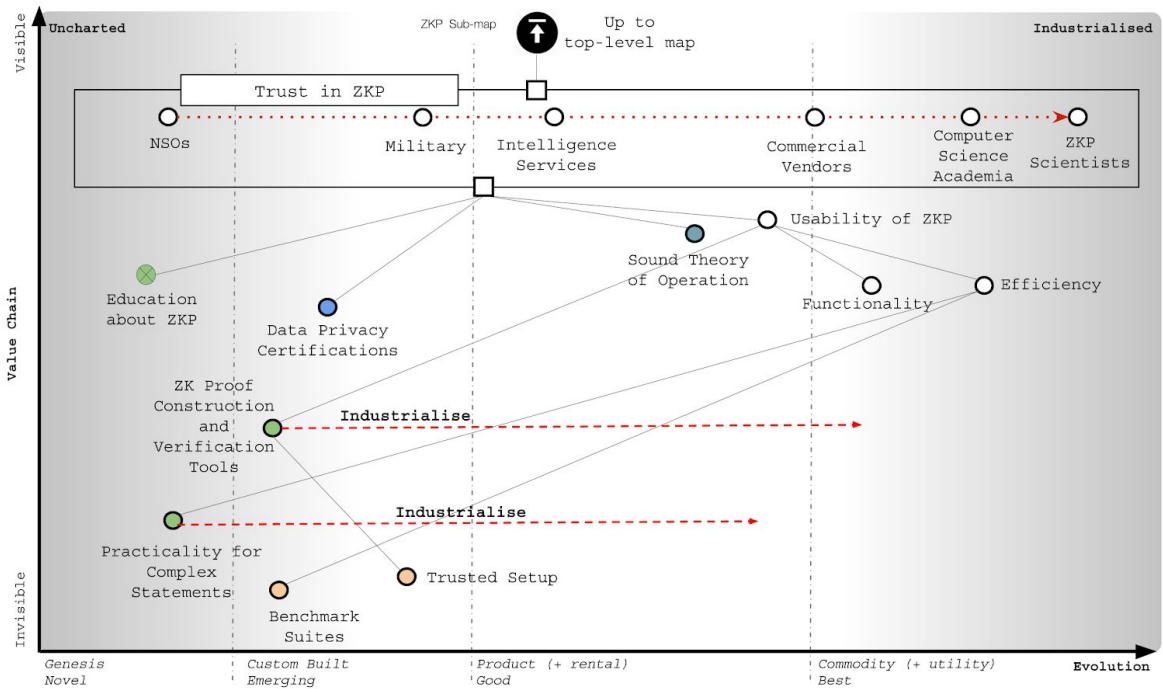

*Figure 10. Wardley map for Zero Knowledge Proofs Technologies.*

Figure 10 presents a Wardley map focused on the details of zero knowledge proofs. While the theory of operation for ZK is at a relatively high state of technology readiness, most of what an end user expects of a computing product is still very early in development.

We need to see improvements in ZK education and privacy certifications to improve NSOs trust in ZK products and services.

# Trusted Execution Environments

## Overview

Trusted Execution Environments (TEEs) provide secure computation capability through a combination of special-purpose hardware in modern processors and software built to use those hardware features. In general, the special-purpose hardware provides a mechanism by which a process can run on a processor without its memory or execution state being visible to any other process on the processor, *even the operating system or other privileged code*. Thus the TEE approach provides Input Privacy.

Computation in a TEE is not performed on data while it remains encrypted. Instead, the execution environment is made secure by the special hardware provided. Such a protected execution environment is often termed an *enclave*. Typically, the memory space of each enclave application is protected from access while resident on the processor chip, and then AES-encrypted when and if it is stored off-chip. Registers and other processor-local state of the enclave are protected from access. Code entry and exit points are tightly controlled, so that execution cannot easily switch between the enclave and the unprotected application that envelops it.

Another significant feature of enclaves is that other processes (whether local or remote) that must trust an enclave can receive *attestation* that the enclave is genuine, and that the code running in it (and in fact the static parts of its memory space) are exactly what is expected. Such attestation is guaranteed using cryptographic capabilities such as digital signatures and hash functions. We note that enclaves can enable Output Privacy and Access Control when the attested code includes specific computations that provide those features.

## Examples of Applied Uses

While other secure computation approaches that protect Input Privacy tend to be slow relative to processing "in the clear" and tend not to scale well with increasing data set size, TEEs often perform and scale well. Relational databases are one application where TEEs are useful because of their performance and scalability. In a typical relational DB application, a data provider might provide an encrypted dataset to a user. Once the user's enclave attested correctly, the data provider might then provide the enclave (over a private channel) with the decryption key for the provided data. The enclave can then internally decrypt the provided data and perform computation as needed. Because TEEs allow for interaction with non-privileged code, interfaces can be provided that allow users to interact with the database application in the same way that users interact with typical relational databases.

Enclave computation can also support *streaming* data applications, where data arrives continuously and is processed through analytics upon arrival. A useful enhancement is that multiple enclaves can be linked together, so that analytics over many data sources can be integrated into one result dataflow without the need to perform all analytics within one enclave. Thus large-scale streaming analysis, such as streaming-rate analytics over sales and shipping data can be accomplished efficiently.

Enclaves also lend themselves to computation "in the small". For example, some server-side banking applications rely on attestation and enclave computation on client-side platforms (such as a user's laptop). For example, a banking server might use a client-side enclave to achieve digital signatures on banking transactions, while protecting the signing key from compromise by any malware that might be running on the laptop.

## Adversary Model and Security Argument

The adversary model most typically used for enclave computing includes a privileged adversary running on the same platform as the enclave, seeking to execute code of the adversary's choice *outside* the enclave in order to access the state of the process running inside the enclave. The security argument against such adversaries is that special-purpose hardware mechanisms prevent any code running outside an enclave from learning any state private to the enclave. Such special-purpose hardware assures for example that virtual memory mapping does not allow processes outside an enclave from mapping physical memory pages also mapped to enclave-private virtual memory. Other hardware features assure that the processor cannot jump into enclave code except at pre-defined legal locations, and that interrupts or other control instructions outside the enclave cannot cause execution from inside to branch to outside code without first securing the enclave and preventing disallowed access.

Another relevant class of adversaries may attempt to inject or replace code running in an enclave in order to allow exfiltration of secrets in the enclave. The security argument against such adversaries combines the protections above and the notion of attestation. Hardware protections assure, for example, that once an enclave is initiated, no change to its code or static data can be achieved from outside the enclave. Once initiated, processes outside the enclave can receive cryptographic attestation that includes a signature of the code inside the enclave, so that the outside process can be assured of all code that can run in the enclave.

## Costs of Using the Technology

Enclave computation is usually comparable in speed to computing "in the clear". Examples that we know of display slowdowns of up to 20% against computing in the clear for relatively small data (on the order of 100MB or so, including application code and data). Other examples show that as data scales towards the Gigabyte range, slowdown may rise to as much as a factor of 6 or 8 times, still far better than MPC or FHE performance.

Use of enclave computation does require the use of specific hardware that includes enclave features. For example, Intel(R) SGX™ features seem to be included in processors of the Skylake™ generation and beyond. Some TEE providers enable virtualization as well, but only virtualization on top of TEE-equipped hardware.

## Availability

Perhaps the most notable enclave capability today is found in Intel(R) processors. Intel's Software Guard Extensions (SGX)™ provide enclave computing in Skylake™ processors and their successors. Virtualization of SGX is an emerging capability, currently (it seems) supported on KVM platforms running on Intel processors. ARM's Trustzone and AMD's Platform Security Processor also offer TEE capability.

The software offerings are diverse as well, ranging from implementation support libraries to privacy-preserving data processing platforms. Some frameworks support the application developer by providing convenience and portability, including Baidu's Rust SGX SDK and MesaTEE, Google's Asylo, Microsoft's Open Enclave SDK, Fortanix's Rust Enclave Development platform and the SGX Linux Kernel Library

Others have more focused applications, like SCONE, a container mechanism with SGX support, R3 Corda for Open Source Blockchain and Sharemind HI, an SGX-powered privacy-preserving analytics platform.

There is also an active research community with projects like Opaque, Ryoan, Graphene Library OS, EnclaveDB, KissDB-SGX and VeritasDB being developed for various secure querying purposes. Projects working on non-analytical goals include SGX Enabled OpenStack Barbican Key Management System and SGX-Log: Securing System Logs With SGX.

Multiple cloud providers offer SGX hardware where one can run these applications when one does not have direct access to such hardware. For example, in Quarter 1 2019, Microsoft has the Azure Confidential Computing program, IBM Cloud provides bare metal machines with SGX support and Alibaba Cloud has SGX machines as well. IBM and Alibaba are also offering secure key management services powered by SGX, by integrating Fortanix offerings.

## Wardley Map

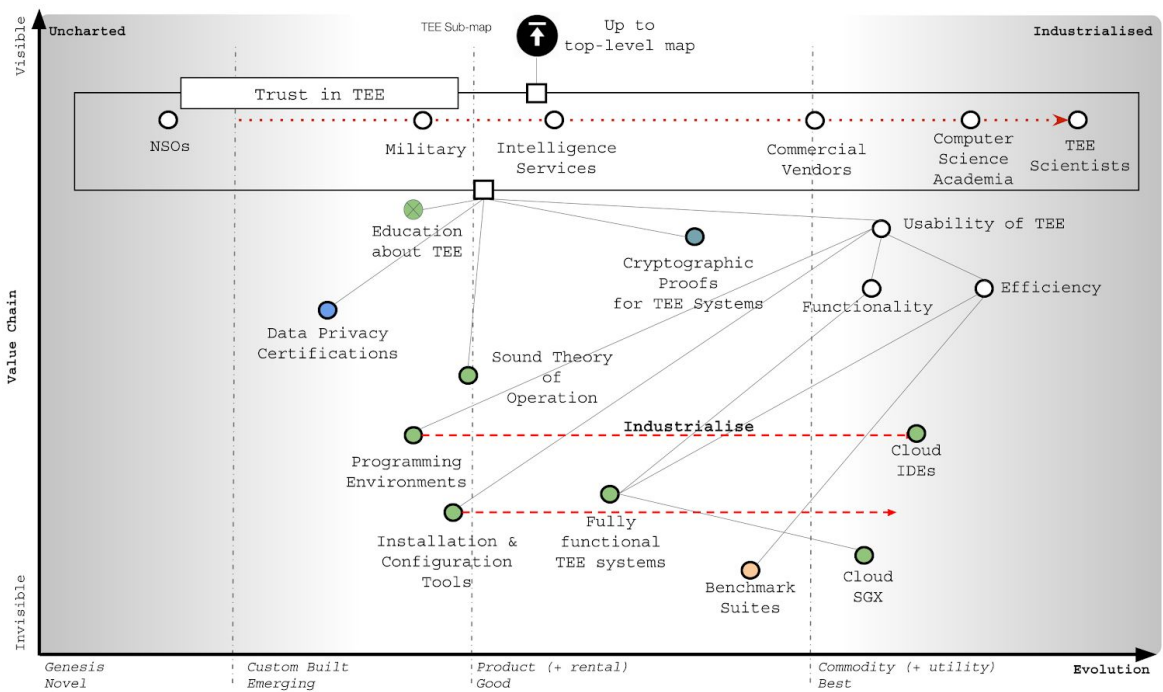

*Figure 11. Wardley map for Trusted Execution Environments.*

Figure 11 presents a Wardley map focused on the details of trusted execution environments. The theory of operation for TEE is at a relatively high state of technology readiness. However, much of what an end user expects in terms of usability of a computing product is still very early in development for TEE. That said, there are emerging products and services that support TEE. Some cloud environments, such as Microsoft Azure and IBM's cloud service offer TEE capability, while others such as Amazon cloud services, do not. The key shortfall at this point in time is the lack of easy to use development environments for TEE, which would enable general programmers to use these capabilities efficiently and configure them correctly. Another current shortfall is that leading TEE's such as Intel SGX require interaction directly with the technology provider in order to properly use these security capabilities.

We need to see improvements in TEE privacy certifications to improve NSOs trust in TEE products and services.

# Standards

## Existing Standards

**ISO/IEC 29101:2013 (Information technology – Security techniques – Privacy architecture framework)** is one of the oldest standards efforts that handles secure computing. It presents architectural views for information systems that process personal data and show how Privacy Enhancing Technologies such as secure computing, but also pseudonymisation, query restrictions and more could be deployed to protect Personally Identifiable Information.

**ISO/IEC 29101** pre-dates the European General Data Protection Regulation (GDPR), so it does not include all the latest knowledge on secure computing and its role in regulation. For example, it is unaware of the view of anonymised processing and using secure computing might actually not be processing in the sense of the law.

**ISO/IEC 19592-1:2016 (Information technology – Security techniques – Secret sharing – Part 1: General)** focuses on the general model of secret sharing and the related terminology. It introduces properties that secret sharing schemes could have, e.g. the homomorphic property that is a key aspect for several MPC systems.

**ISO/IEC 19592-2:2017 (Information technology – Security techniques – Secret sharing – Part 2: Fundamental mechanisms)** introduces specific schemes. It starts with the classic ones like Shamir and replicated secret sharing. All schemes are systematically described using the terms and properties from Part 1. There were originally plans to have more parts for this standard that would describe MPC paradigms, but work has not started yet.

## Standards In Development

**ISO/IEC 18033-6 (Information technology security techniques – Encryption algorithms – Part 6: Homomorphic encryption)** is a standard on homomorphic encryption schemes

Given the more conservative nature of ISO/IEC when it comes to encryption schemes, it is attempting to focus on the ones with multiple known industrial uses. However, as it is still work in progress, it is unclear how it will turn out in the end.

The **Homomorphic Encryption Standardization Initiative**[22] is an open standardisation initiative for fully homomorphic encryption with participants from industry, government and academia. The initiative attempts to build broad community agreement on security levels, encryption parameters, encryption schemes, core library API, and eventually the programming model, with the goal of driving adoption of this technology.

**ISO/IEC 20889 (Privacy enhancing data de-identification terminology and classification of techniques)** is another project that approaches privacy technologies a bit differently. This project will result in a standard that describes ways to turn identifiable data into de-identified data. Here, the choices include various noise-based techniques, cryptographic techniques and more.

# UN Global Platform Marketplace

The UN Global Working Group on Big Data launched the UN Global Platform Marketplace in early May '19 at the 50th United Nations Statistical Commission. The Marketplace provides a central place to search for trusted algorithms, methods, learning, services and partners.

We have provided a searchable directory for methods, algorithms, learning and partners related to privacy preserving techniques,
https://marketplace.officialstatistics.org/learnings?statistics_area=337

---

[22] Online website: http://HomomorphicEncryption.org (last accessed July 2nd, 2018)

UN Handbook on Privacy-Preserving Computation Techniques        46

# Legal / Legislation

## Legal Research on Secure Multiparty Computation

There have been multiple efforts to analyse the relations of secure multiparty computation and data protection regulations. Below, you'll find some notable results.

One of the first significant precedents for secure multiparty computation was reached in Estonia with the Private Statistics project in 2015[23]. In the project, 10 million identifiable tax records were linked with 600 000 identifiable education records and statistically analysed using secure multiparty computation. The Data Protection Agency, after studying the technical and organisational controls of the system, stated that no personal data was processed. The precedent has also been upheld with the MPC servers hosted in the public cloud[24].

The PRACTICE project (European Commission Framework Programme 7) spent significant effort in analysing legal aspects of secure computing technologies. The report[25] studies the Estonian precedent described above under the European General Data Protection (GDPR) regulation and finds that precedent can be upheld under the GDPR[26].

Further research has been performed by the SafeCloud project[27] and SODA project[28]

---

[23] Dan Bogdanov, Liina Kamm, Baldur Kubo, Reimo Rebane, Ville Sokk, Riivo Talviste. **Students and Taxes: a Privacy-Preserving Social Study Using Secure Computation**. In Proceedings on Privacy Enhancing Technologies, PoPETs, 2016 (3), pp 117–135, 2016. http://dx.doi.org/10.1515/popets-2016-0019

[24] **National Special Education Data Analysed Securely**.
https://sharemind.cyber.ee/national-special-education-data-analysed-securely/ (Last accessed July 19th, 2018)

[25] **Evaluation and integration and final report on legal aspects of data protection**. PRACTICE project deliverable D31.3.
https://practice-project.eu/downloads/publications/year3/D31.3-Evaluation-and-integration-and-final-report-on-PU-M36.pdf

[26] Prof. Dr. Gerald Spindler, Philipp Schmechel. **Personal Data and Encryption in the European General Data Protection Regulation.** 7 (2016) JIPITEC 163 para 1. http://www.jipitec.eu/issues/jipitec-7-2-2016/4440

[27] The SafeCloud project. http://www.safecloud-project.eu/results/deliverables See deliverable D2.3. Last accessed July 19th, 2018.

[28] The SODA project. https://www.soda-project.eu/deliverables/ See deliverable D3.1. Last accessed July 19th, 2018.

UN Handbook on Privacy-Preserving Computation Techniques     47

## Legal Research on Other Proposed Technologies

At the time of writing this handbook, the authors were not aware of validations done to other privacy-preserving technologies that would be on the same level as what has been done with multi-party computation. There has been work towards that end, with a case study on differential privacy[29]. The task team will be monitoring developments in this area.

## Recurring Events and Forums on Secure Computation

| Name | Description |
| --- | --- |
| Theory and Practice of Multi-Party Computation Workshop | The TPMPC workshops aims to bring together practitioners and theorists working in multi-party computation. <br><br> The TPMPC workshops continue a tradition of workshops started in Aarhus, Denmark in 2012. <br><br> See http://www.multipartycomputation.com/ for details on the next workshop. |
| RSA Conference | RSA Conference conducts information security events around the globe that connect you to industry leaders and highly relevant information. They deliver, on a regular basis, insights via blogs, webcasts, newsletters and more so you can stay ahead of cyber threats. <br><br> See https://www.rsaconference.com/ for a list of events. |
| Real World Crypto | Real World Crypto Symposium aims to bring together cryptography researchers with developers implementing cryptography in real-world systems. The conference goal is to strengthen the dialogue between these two communities. Topics covered focus on uses of cryptography in real-world environments such as the Internet, the cloud, and embedded devices. <br><br> See https://rwc.iacr.org/ for details on the next conference. |
| Theory and Practice of Differential Privacy | The overall goal of TPDP is to stimulate the discussion on the relevance of differentially private data analyses in |

---

[29] Nissim, Kobbi, Aaron Bembenek, Alexandra Wood, Mark Bun, Marco Gaboardi, Urs Gasser, David O'Brien, Thomas Steinke, and Salil Vadhan. **Bridging the gap between computer science and legal approaches to privacy**. Harvard Journal of Law & Technology. Volume 31, Number 2 Spring 2018.

| | practice. TPDP is a recurring workshop co-located with different conferences in security or machine learning.<br><br>Search the web for "Theory and Practice of Differential Privacy" for details on the workshop |
|---|---|

## Training

| Course | Description | |
|---|---|---|
| Secure Computation | Secure Computation course offered by Indian Institute of Science covering secret sharing schemes, oblivious transfer to impossibility results and zero-knowledge proofs. | |
| Secure Multi-Party Computation at Scale | Boston University course that covers mathematical and algorithmic foundations of MPC, with an additional focus on deployment of state-of-the-art MPC technologies. | |
| Bar-Ilan 1st Winter School on Secure Computation and Efficiency | Graduate level course on MPC, mainly concentrating on the theory | https://www.youtube.com/watch?v=z3U-5mf6hGw&list=PL8Vt-7cSFnw2rc1Y6qBSFbFbgOIWsOlsV |
| Bar-Ilan 5th Winter School on Practical Advances in Multi-Party Computation | Graduate level course on MPC, focusing more on practical algorithms | https://www.youtube.com/watch?v=C6WRWtym2JY&list=PL8Vt-7cSFnw00U0jMSgAZJrpIKG-m_0gH |
| Bar-Ilan 7th Winter School on Differential Privacy | Graduate level course on Differential Privacy | https://www.youtube.com/playlist?list=PL8Vt-7cSFnw1Ii73YXZdTaiAeXFkmWWRh |
| A short tutorial on differential privacy. Speaker: Borja Balle (Amazon Research, UK). | The first half of the tutorial will introduce the basic ideas and provide a brief survey of some of their applications in | https://www.youtube.com/watch?v=ZUsW_4GdEK8 |

UN Handbook on Privacy-Preserving Computation Techniques                                                                 49

| | | |
|---|---|---|
| | privacy-preserving machine learning.<br><br>In the second half of the tutorial we will present several variants of the original definition of differential privacy, and discuss the roles each of these definitions plays in practical applications.<br><br>This is the first one of a series of talks in the context of the interest group on [Privacy-Preserving Data Analysis](#) | |
| Zero Knowledge Proofs | Proceedings of the continuing workshop to standardize zero knowledge proofs, and a site that overviews existing implementations of zero knowledge protocols | [https://zkproof.org](https://zkproof.org)<br><br>[https://zkp.science](https://zkp.science) |

UN Handbook on Privacy-Preserving Computation Techniques    50